# Properties of Type III and Type IIIb Bursts in the Frequency Band of 8 - 80 MHz during PSP Perihelion at the Beginning of April 2019

V.N.Melnik (1), A.I. Brazhenko (2), A.A. Konovalenko (1), A.V. Frantsuzenko (2), S.M. Yerin (1), V.V. Dorovskyy (1), I.M. Bubnov (1)

(1) Institute of Radio Astronomy, National Academy of Sciences of Ukraine, Kharkov, Ukraine
(2) Gravimetrical Observatory, Poltava, Ukraine

**Abstract**

Abstract Properties of type III and type IIIb bursts in the frequency band of 8 - 80 MHz observed by the radio telescopes Ukrainian Radio Interferometer of NASU-2 (URAN-2) (Poltava) and Giant Ukrainian Radio Telescope (GURT) (Kharkiv) during the Parker Solar Probe (PSP) perihelion in April 2019 are discussed. These correspond of those that were observed by PSP at frequencies <19 MHz. We analyze dependences of drift rates and durations on frequency for these bursts. We show that drift rate dependences on frequency agree well with those derived from the Newkirk corona if source velocities are between 0.17 and 0.2 c for both type IIIb bursts and type III bursts under the assumption that the first ones are fundamentals and the second ones are their harmonics. However, all observational dependences are flatter in comparison with the dependences for a Newkirk corona. We assume that this can be related with coronal temperature decreasing at heliocentric distances from 1.35 to 6.5 solar radii. Duration dependencies of type IIIb and type III bursts on frequency in the range of 10 - 70 MHz are also obtained. We note that the durations and drift rates of type III bursts as well as their dependences on frequency differ essentially from those for type IIIb bursts.

1. **Introduction**

Type III bursts are those that drift from high frequencies to low ones, as a rule, with decreasing drift rates (Suzuki and Dulk, 1985). Already in the first paper devoted to type III bursts, Wild (1950) estimated the velocity of type III source as 0.3 c (c being the speed of light). Further it was established that the source velocity could change in the range from 0.2 to 0.6 c (Zaitsev, Mityakov, and Rapoport, 1972). Observational drift rate dependences on frequency were analyzed many times (Alvarez and Haddock, 1973b; Mann et al., 1999; Melnik et al., 2011; Zhang, Wang, and Ye, 2018); however, the one most debated was that of Alvarez and Haddock (1973b) $df/dt = -0.01 f^{1.84}$, which was obtained in the wide frequency range of 75 kHz to 550 MHz on a large observational data set with various radio telescopes in different years. Recently, Zhang, Wang, and Ye (2018) analyzed the drift rate dependence for type III bursts in the frequency range of 10 - 80 MHz using observations by the NDA (Nancay Decameter Array) during the period 2012 - 2017. They processed 1389 type III bursts and found the dependence $df/dt = -0.0672 f^{1.23}$ with large dispersion of drift rates at different frequencies. Such dispersion might be connected with three causes:

- different type III bursts are generated by electron beams with different velocities;
- analyzed bursts were observed at different times thus above different active regions and as a consequence were generated in different plasmas;
- the authors did not separate the fundamental type III bursts from their harmonics therefore radio emission at the same frequency might have been generated at different places and as a consequence drift rates could differ essentially.

In such case seeking for a universal dependence is not correct. In our opinion analysis of drift rate dependences for the same burst in the wide frequency band separating the fundamental and harmonic components seems to be more adequate. Nowadays such opportunity exists. In fact there are some sensitive radio telescopes such as LOw Frequency ARray (LOFAR), Long Wavelength Array (LWA), Ukrainian T-shaped Radio telescope - 2 (UTR-2), Ukrainian Radio interferometer of NASU -2 (URAN-2), New extension in Nancay upgrading LOFAR (NenuFAR), Giant Ukrainian Radio Telescope (GURT) with high frequency-time resolution, which work in the wide frequency band >8 MHz along with the Parker Solar Probe (PSP) mission which can provide observations at frequencies <19 MHz. Study of individual type III bursts will allow to develop more appropriate models for the solar corona up to some solar radii above active regions. Also this will allow to find type III source velocities for different events and to understand in what way they vary while propagating in the solar corona.These studies will as well help to clear up details of propagation of electron beams responsible for type III bursts (Zaitsev, Mityakov, and Rapoport, 1972; Grognard, 1985; Muschietti, 1990; Mel'Nik, Lapshin, and Kontar, 1999; Reid and Kontar, 2018a,b) and processes of radio emission at the first and the second harmonics (Ginzburg and Zhelezniakov, 1958; Smith, Goldstein, and Papadopoulos,1976; Mel'Nik and Kontar, 2003).

The study of the dependence of type III burst duration on frequency aims to answer the question on in what way both the electron beams and the radio waves propagate in the coronal plasma (Suzuki and Dulk, 1985; Rutkevych and Melnik, 2012). The influence of the effect of electromagnetic wave scattering in the coronal plasma on the durations and apparent sizes of type III bursts has been lately widely discussed (Kontar et al., 2017; Sharykin, Kontar, and Kuznetsov, 2018; Kontar et al., 2019; Krupar et al., 2020). It was pointed out that the duration of the fundamental and harmonic components in type IIIb-III pairs differed substantially in the frequency band of 10 - 30 MHz (Melnik et al., 2018). Studies extending the frequency band above 30 MHz (LOFAR, LWA, NenuFAR, GURT) and below 10 MHz (PSP) appear to be promising for clarifying this point.

In this article type III and type IIIb bursts in the frequency band of 8 – 80 MHz, which were observed by the radio telescopes URAN-2 and GURT on 6, 8, and 9 April 2019 during the PSP perihelion, are discussed. Type III and type IIIb bursts were initiated by electrons accelerated above the active region NOAA 2738 near the eastern limb according to Krupar et al. (2020). Frequency dependences of burst drift rates and durations were found in the frequency band of 8 – 80 MHz. We show that the observed drift rate dependences can be understood in the context of Newkirk model (Newkirk, 1961) of the coronal plasma if electron beam velocities are about 0.2 c both for fundamental and harmonic components. We also show that the duration difference between type III and type IIIb bursts is essential in the frequency range of 8 - 80 MHz. Apparently this fact cannot be explained by scattering effects on coronal inhomogeneities.

**2. Observations**

At the beginning of April 2019 solar observations were carried out by Ukrainian radio telescopes URAN-2 (Poltava) and GURT (Kharkiv) (Konovalenko et al., 2016). The radio telescope URAN-2 works in the frequency range of 8 - 33 MHz (Brazhenko et al., 2005). Its effective area is 28,000 m2. URAN-2 is calibrated and its sensitivity is about 500 Jy. It can measure polarization in the frequency band of 8 - 33 MHz. The frequency-time resolution in these observations was 4 kHz - 100 ms (Zakharenko et al., 2016).

The working frequency band of the radio telescope GURT subarray is 8 – 80 MHz (Konovalenko et al., 2016). Its effective area is 350 m2 and the frequency-time resolution is 9 kHz - 50 ms. Currently the calibration is absent. Both radio telescopes have a dynamic range of 90 dB.

Some type III bursts observed at the beginning of April 2019 were analyzed by Krupar et al. (2020) in the frequency range of 10 kHz - 19 MHz. By that time PSP was at distances ranging from $37.5 R_\odot$ to $53.8 R_\odot$ and its longitudinal angle was about $68°$ east from the Sun-Earth line. Observations were carried out with the radio telescopes URAN-2 and GURT in the frequency range of 8 - 80 MHz. Krupar et al. (2020) found that the electron beams, which generated these type III bursts, were accelerated in active region NOAA 2738 situated on the eastern limb of the Sun (Figure 1). We will discuss properties of type III and type IIIb bursts observed on 6, 8, and 9 April 2019. Additionally we will investigate drift rate dependences on frequency for these bursts, as well as their duration dependence on frequency.

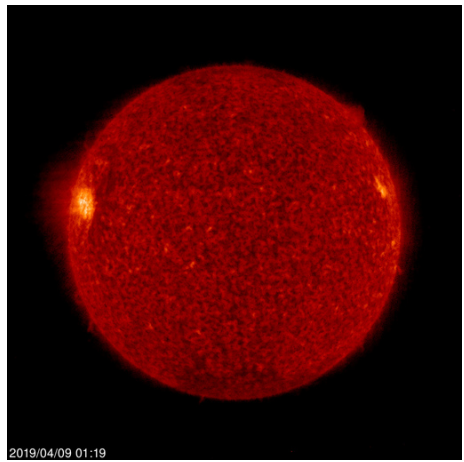

**Figure 1.** Active region NOAA 2738 on 9 April 2019 according to SOlar Heliospheric Observatory (SOHO).

### 2.1. Type III and Type IIIb Bursts at 11:07:30 UT on 6 April 2019

Dynamic spectra of type III and type IIIb bursts according to URAN-2 and GURT are presented in Figure 2a-b. PSP also observed this type III burst about 7 minutes earlier (Figure 2c). According to URAN-2 the flux of the type III burst equalled 13 s.f.u. at a frequency of 33 MHz and it increased towards low frequencies, reaching about 70 s.f.u at 9 MHz. Polarization of this burst was not high, actually not exceeding 20%. The low polarization is typical for radio emission at the second harmonic (Suzuki and Dulk, 1985). There was a type IIIb burst registered mainly at frequencies 36 - 70 MHz and some of its stria at frequencies <30 MHz before the burst. The frequency ratio of type III and type IIIb bursts at the same time varied from 1.83 to 1.99.
As an example Figure 3 shows profiles of type III and type IIIb bursts at frequencies 74.16 and 37.41 MHz as well as at frequencies 49.19 and 24.67 MHz. We see that maxima of type IIIb bursts were observed at the same time as those of type III bursts at higher frequencies. Frequency ratios for these pairs are 1.98 and 1.99 correspondingly. This observation favors the harmonic phenomenon of

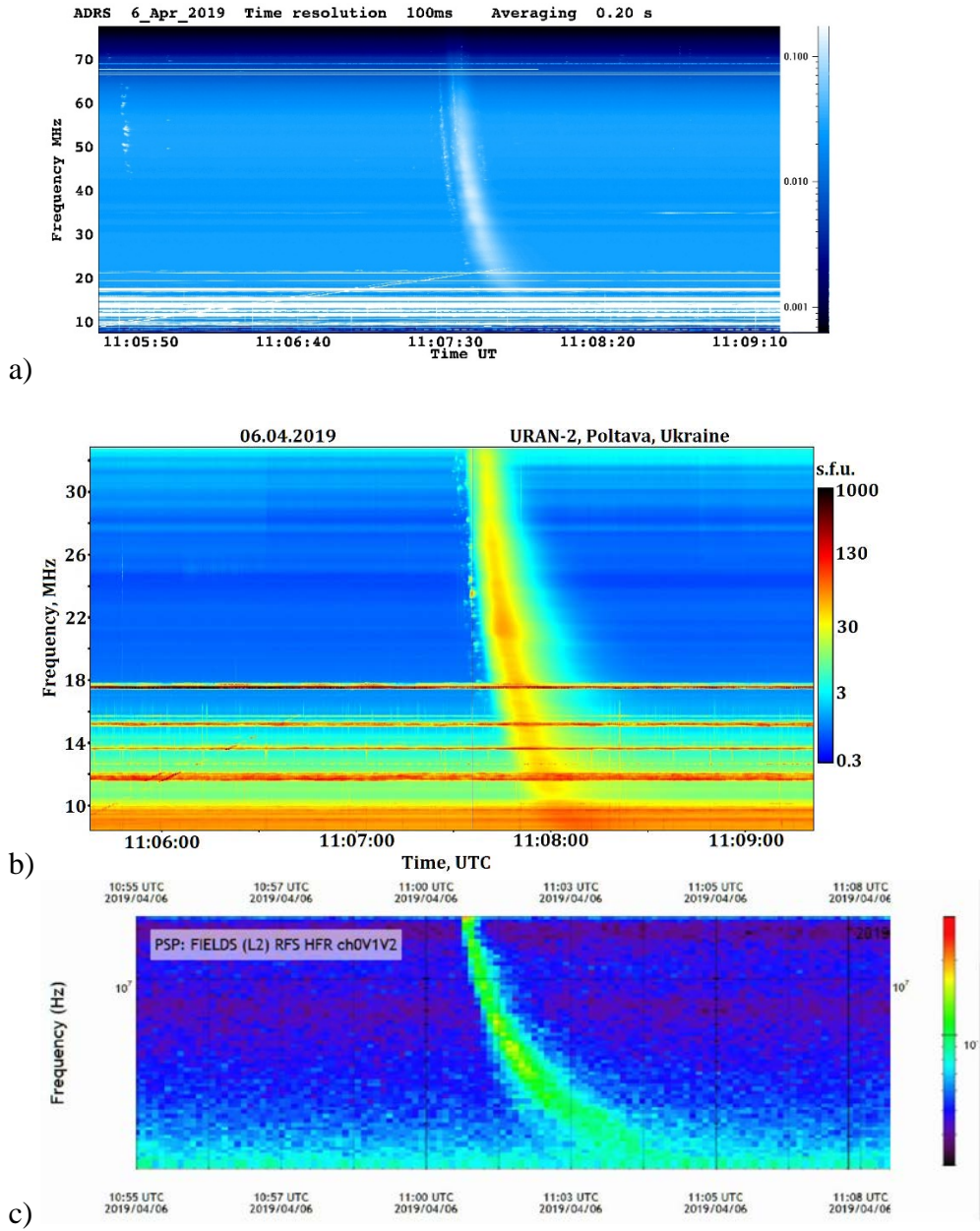

a)

b)

c)

**Figure 2.** Dynamic spectra of type III and type IIIb bursts at 11:07:30 UT on 6 April 2019 according to GURT (a) and URAN-2 (b). The type III burst observed by PSP is also shown (c).

As an example Figure 3 shows profiles of type III and type IIIb bursts at frequencies 74.16 and 37.41 MHz as well as at frequencies 49.19 and 24.67 MHz. We see that maxima of type IIIb bursts were observed at the same time as those of type III bursts at higher frequencies. Frequency ratios for these pairs are 1.98 and 1.99 correspondingly. This observation favors the harmonic phenomenon of radio emission of IIIb-III pairs (Melnik et al., 2018), namely that type IIIb bursts are radiated at the first harmonic and type III bursts at the second harmonic. In Figure 4 the frequency dependence of the drift rate for a type III burst observed at 11:07:30 UT in the range 10 - 74 MHz is shown together with that for a type IIIb burst observed in the frequency range 23-70 MHz. Errors on the drift rates of type III bursts are connected with fluctuations of maximum time and in most cases are not higher than 10%. For type IIIb bursts errors are determined by the dispersion of stria-bursts on frequency and time and they reach 30%. Figure 4 also shows the drift rate dependence on frequency for Newkirk coronal plasma according to the equation

$$df/dt = (f/2)(1/n)(dn/dr)V \qquad (1)$$

where $n$ is the density in the Newkirk corona $n(r) = 4.2 \cdot 10^4 \cdot 10^{4.32 R_\odot / r}$ ($R_\odot$ is the solar radius) (Newkirk, 1961) and $V = 0.17c$ is the speed of the electron beam responsible for type III burst if radio emission happened at the second harmonic.

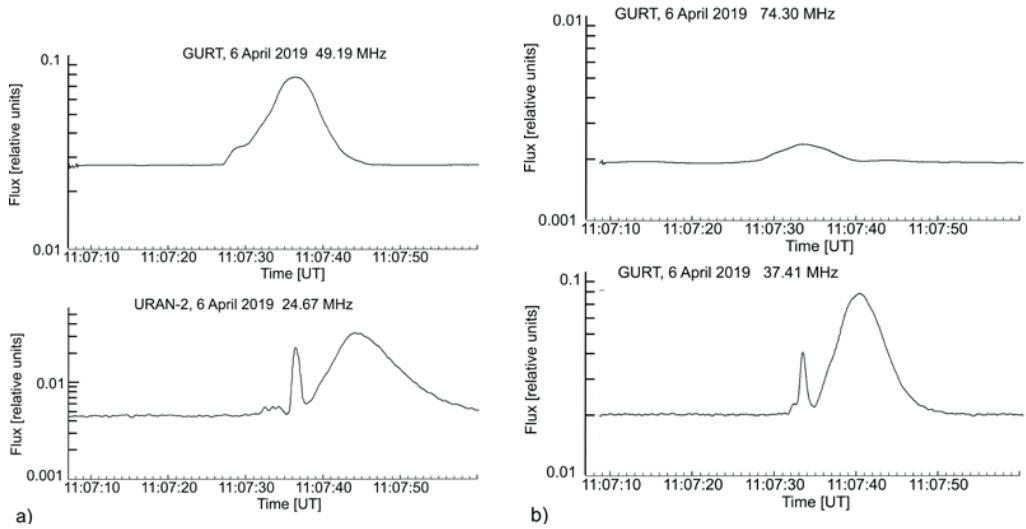

**Figure 3.** Profiles of type IIIb and type III bursts at frequencies 24.67 and 49.19 MHz (a) and at frequencies 37.41 and 74.30 MHz (b) for the event on 6 April 2019.

It is seen that the data lie a little lower than calculated at higher frequencies and a little higher at lower frequencies. Observational dependence is approximated by the equation $df/dt = -2.07(f/30MHz)^{1.63}$ ($f$ in MHz and $df/dt$ in MHz/s). For the type IIIb burst, such dependence is flatter $df/dt = -3.85(f/30MHz)^{1.43}$, however situated higher as it should if type IIIb burst was fundamental (Melnik et al., 2018). Considered that way and with a source speed $V = 0.17c$ in the Newkirk model the functional dependence is closer to the observational one.

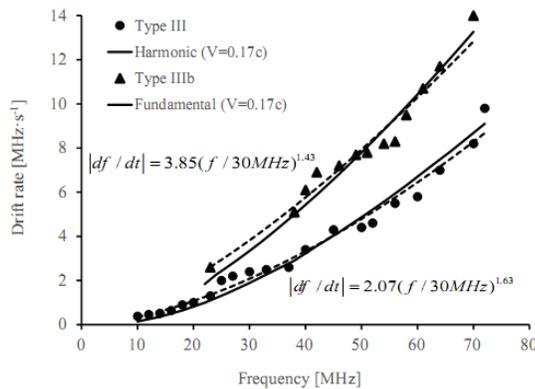

**Figure 4.** Dependences of drift rate on frequency (dashed lines) for the type III burst at 11:07:30 UT and for the type IIIb burst according to URAN-2 and GURT. Frequency dependences for drift rates

in the Newkirk model for a source speed 0.17 c are shown for fundamental and harmonic by solid curves.

Durations for type III and type IIIb bursts observed by URAN-2 and GURT are presented in Figure 5. The duration of the type III burst was measured in a 1 MHz interval. Because the shapes of time profiles of both type III burst and type IIIb burst are close to Gaussian distributions errors on the measured durations were not higher than a few percent. Even at close frequencies, durations can differ which seems to be connected with the conditions for the generation of type III bursts. On the other hand, durations of type IIIb burst, which consists of stria-bursts (for this particular type IIIb burst their number was 32), can be different essentially for neighboring bursts. As an example of such cases see the durations of stria-bursts at 45 MHz (Figure 5). Nevertheless there is some frequency dependence which can be approximated by the equation $D = 1.12(f/30MHz)^{-0.5}$ ( $f$ in MHz and $D$ in seconds ). For type III burst duration we found the dependence $D = 7(f/30MHz)^{-0.5}$. We see that durations of type III burst at all frequencies are longer than those of type IIIb burst by a factor of about 6. Such difference was already indicated earlier for the frequency band 10 - 30 MHz (Melnik et al., 2018).

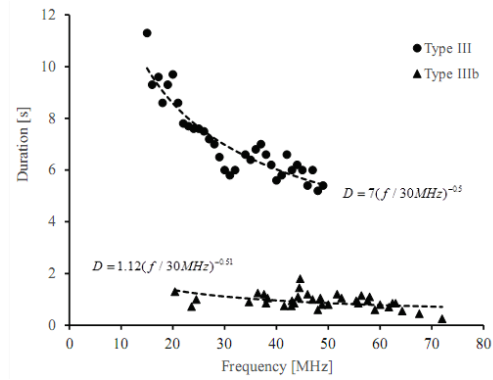

**Figure 5.** Durations of type III and type IIIb bursts at different frequencies.

## 2.2. Type III Burst at 9:47 UT on 8 April 2019

This burst (Figure 6) was also weak, its flux was about 13 s.f.u. at 32 MHz and remained practically the same at all frequencies down to 10 MHz. PSP did not register it due to such low flux. Polarization of this burst was about 10% at all frequencies. So we can suppose that this type III burst was also the harmonic. Before this burst we can see a type IIIb burst at frequencies <30 MHz. Its polarization was high, up to 60% and it seems to be the fundamental. Unfortunately its flux was only a few s.f.u. what made accurate measurements of drift rates and durations difficult. That is why we do not mention those results. The drift rate of the type III burst dependence on frequency in the range of 10 - 74 MHz is shown in Figure 7. It is approximated by the equation $df/dt = -2.86(f/30MHz)^{1.31}$ and is flatter than that for the type III burst on 6 April. This dependence is in a good agreement with the theoretical equation for the drift rate of the harmonic in the Newkirk model with an electron beam velocity $V = 0.2c$. As for the type III burst on 6 April the empirical dependence is flatter than the theoretical one.

Duration of this burst could be measured in the frequency band of URAN-2 because of its high sensitivity. At higher frequencies >33 MHz it was difficult to measure its duration due to lower sensitivity of GURT. In the frequency range 14 - 33 MHz the duration dependence on frequency for this burst is presented in Figure 8 as $D = 5.7(f/30MHz)^{-1.21}$. This dependence is steeper than that for the type III burst on 6 April in the frequency range of 14 - 50 MHz.

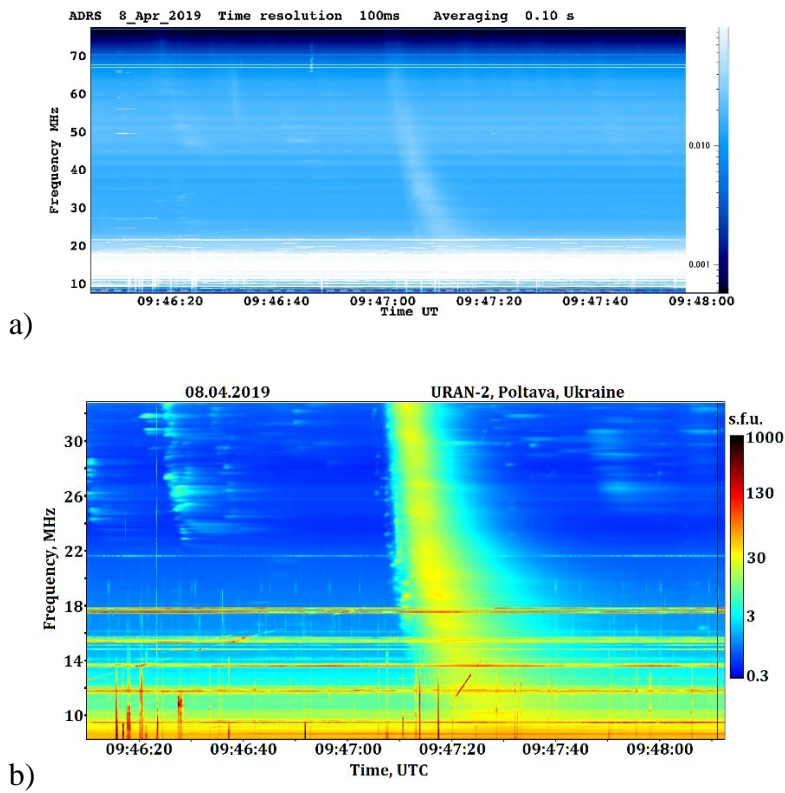

**Figure 6.** Dynamic spectra of the type III burst at 9:47 UT on 8 April 2019 according to GURT (a) and URAN -2 (b) observations.

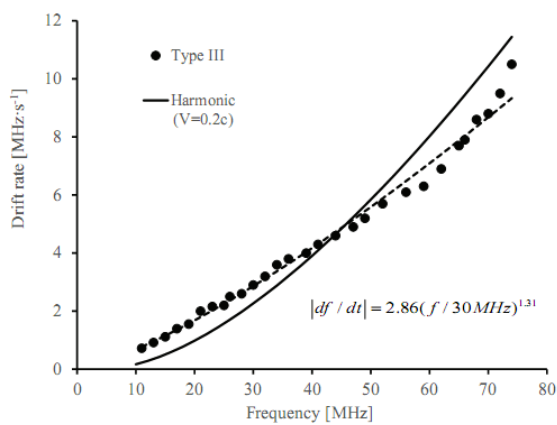

**Figure 7.** Frequency dependence of drift rate for the type III burst observed at 9:47 UT on 8 April 2019.

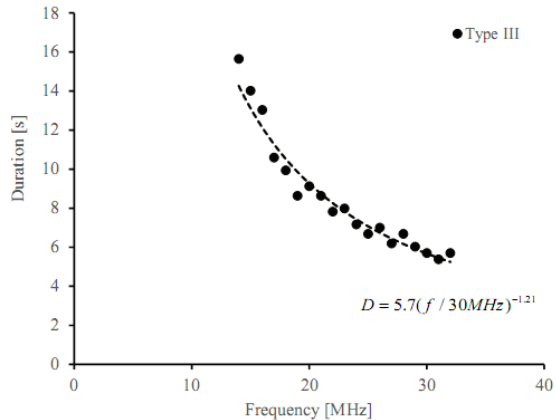

**Figure 8.** Frequency dependence of the duration of the type III burst at 9:47 UT on 8 April in the frequency range of 14-33 MHz.

## 2.3. Type III and Type IIIb Bursts at 12:45 UT on 9 April 2019

Type III and type IIIb bursts shown in Figure 9a-b appeared to be more powerful. This allowed to analyze their properties in a wider frequency range from 14 to 74 MHz. At 32 MHz the type III burst flux was 32 s.f.u. and at 10 MHz its flux increased up to 350 s.f.u. PSP registered this type III burst at lower frequencies (Figure 9c). Its polarization did not exceed 10%. The type IIIb burst had a ux of about 50 s.f.u. at 32 MHz and about 200 s.f.u. at 16 MHz. At frequencies > 33 MHz the flux of the type IIIb was several times higher than that of the type III burst. Nevertheless at frequencies above 50 MHz the type IIIb burst was not visible (Figure 9a). Thus frequency dependences on drift rate and duration could not be obtained. The polarization of the type IIIb burst reached 80%. The frequency ratio for the type III and type IIIb bursts at the same time varied from 1.78 to 2 at different frequencies. It means that this pair of bursts was the harmonic (type III burst) and fundamental (type IIIb burst) components, which is also confirmed by the polarization.

Observational drift rate dependences for type III and type IIIb bursts are presented in Figure 10. They can be approximated by equations $df/dt = -2.61(f/30MHz)^{1.49}$ and $df/dt = -3.47(f/30MHz)^{1.42}$, respectively. We see that the theoretical dependences using the Newkirk model for the harmonic and fundamental with electron beam velocities $V = 0.19c$ and $V = 0.2c$ are in good correspondence with the observational dependences. As for events on 6 and 8 April this observational dependences were flatter than those for theoretical ones. It is worth noting that for this event the drift rates of the type IIIb are 1.5 - 2 times higher than for the type III as pointed out already in Melnik et al. (2018). This feature is natural for harmonic pairs.

Due to insufficient sensitivity of the single GURT subarray we could not measure correctly the durations of the type III burst at frequencies higher than 33 MHz. However, in the frequency range 10 - 33 MHz the dependence of duration on frequency is as follows $D = 6.7(f/30MHz)^{-0.96}$ (Figure 11). Durations of type IIIb bursts were measured in a wider frequency range from 10 MHz to 50 MHz and could be approximated by equation $D = 0.86(f/30MHz)^{-0.49}$. Again as for the event of 6 April durations of the type IIIb burst at different frequencies were actually durations of individual stria-bursts (their total number was 56), and again as previously durations of different individual striae can vary almost twice at the same frequencies. For this IIIb-III pair the durations of the type III bursts exceed those of type IIIb bursts by approximately 8 times. Note also that the power law indexes for these two burst dependences differ considerably.

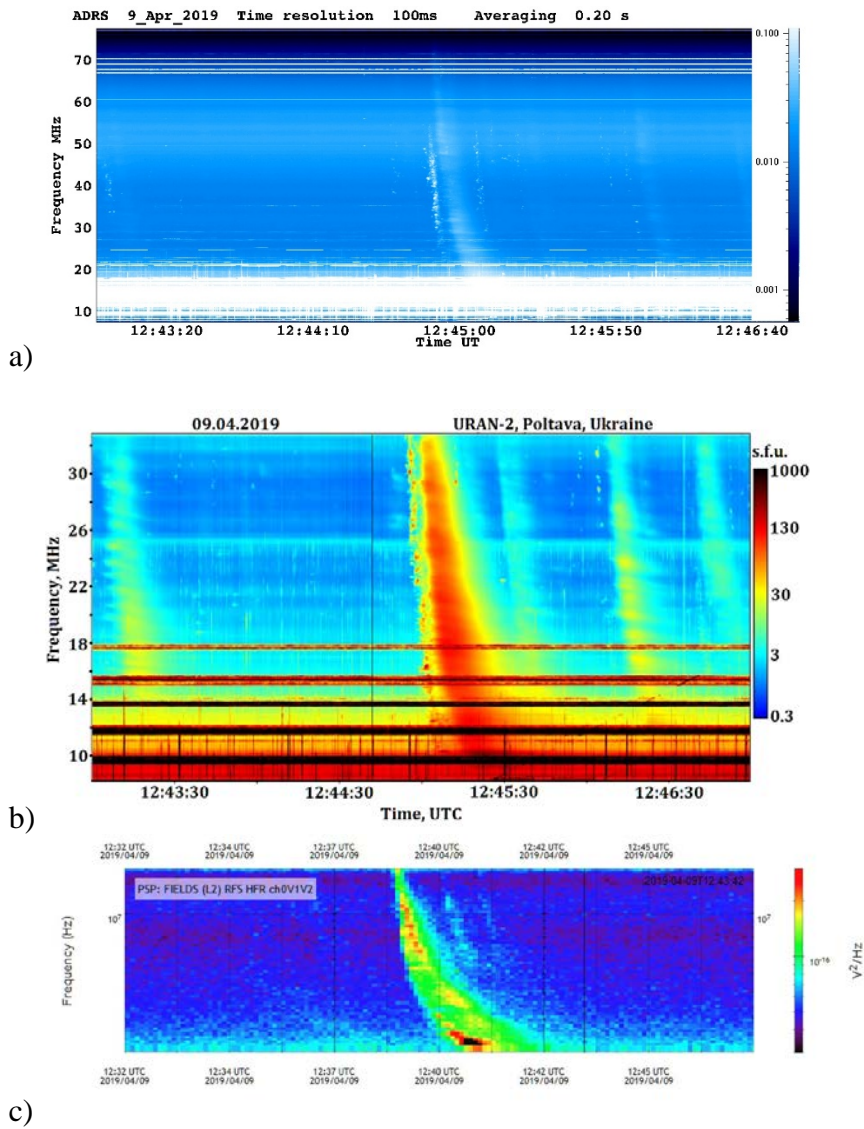

a)

b)

c)

**Figure 9.** Dynamic spectra of type III and type IIIb bursts at 12:45 UT on 9 April 2019 observed by GURT (a) and URAN-2 (b). The same type III burst was observed by PSP about 7 minutes earlier (c).

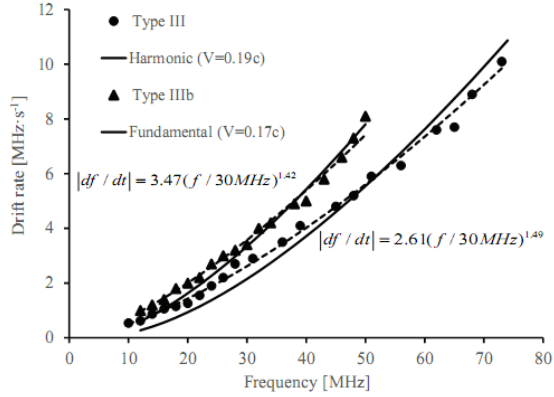

**Figure 10.** Drift rate dependences on frequency (dashed lines) for type III (a) and type IIIb (b) bursts at 12:45 UT on 9 April 2019. Drift rate dependences for fundamental and harmonic for the Newkirk corona are shown by solid lines.

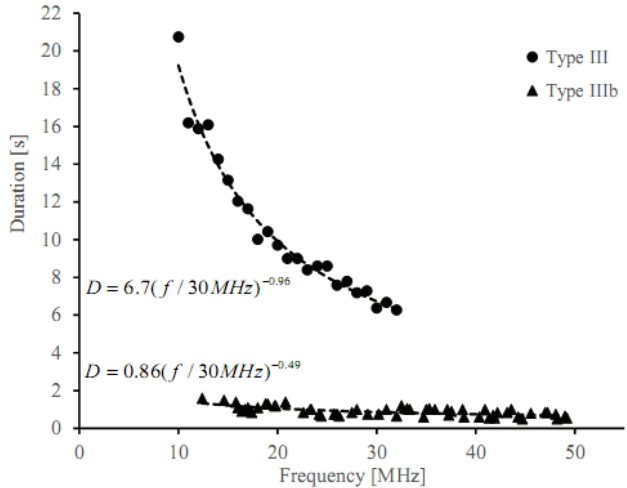

**Figure 11.** Durations for type III and type IIIb bursts at 12:45 UT on 9 April 2019.

## 3. Discussion

### 3.1. Drift Rates of Type III and Type IIIb Bursts

It is widely thought that the drift rate of the type III burst depends on frequency according to the equation $df/dt = -0.01 f^{1.84}$ (in our representation $df/dt = -5.22(f/30MHz)^{1.84}$) (Alvarez and Haddock, 1973b) in the wide frequency band from 75 kHz to 550 MHz. Analyzing type III bursts observed in the frequency range 10 - 80 MHz by the NDA radio telescope in 2012 - 2017 Zhang, Wang, and Ye (2018) found another dependence $df/dt = -0.0672 f^{1.23}$ (in our representation $df/dt = -4.41(f/30MHz)^{1.23}$). One of the features of these data was a very high dispersion of drift rate values (up to several times) at the same frequency. Figure 12 shows the region, indicated with a solid line, in which all drift rates measured by Zhang, Wang, and Ye (2018) were located. This figure also shows drift rates dependences on frequencies for the fundamental and harmonic with electron beams velocities 0.2, 0.3, and 0.6 c for the Newkirk coronal plasma. It is seen that practically all the indicated region covers the area limited by curves for the fundamental with source velocity 0.6 c and the harmonic with velocity 0.2 c. Therefore in our opinion the dispersion in drift rates obtained by Zhang, Wang, and Ye (2018) resulted from merging the fundamental and harmonic radio emissions of type III bursts and the possible different source velocities of bursts. Besides, those type III bursts were observed in different conditions above different active regions meaning that coronal plasmas could differ from the standard Newkirk model in general which can also lead to the large dispersion of drift rates.

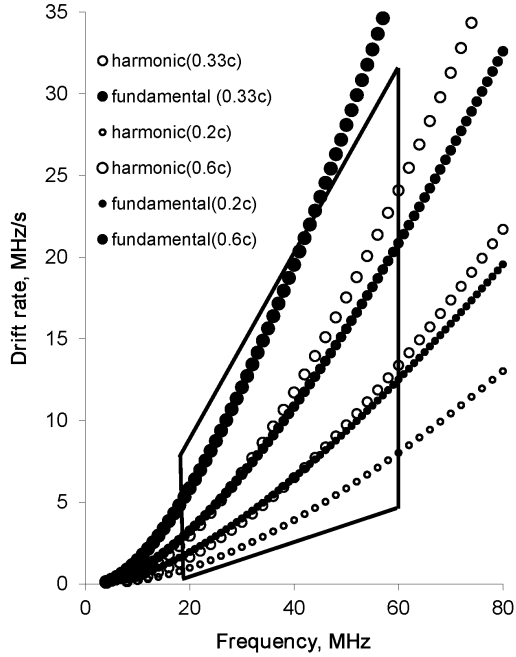

**Figure 12.** Dependences of drift rates for type III bursts in the Newkirk model (Newkirk, 1961) in the case of fundamental and harmonic for electron beam velocities 0.2, 0.3, and 0.6 c. The region of measured drift rates at different frequencies for type III bursts observed by NDA (Zhang et al., 2018) is outlined by solid line.

Drift rates dependences for type III and type IIIb bursts obtained in this article are in good agreement with dependences for the Newkirk model (Newkirk, 1961) for source speeds 0.17 - 0.2 c both for the fundamental and harmonic and do not contradict Dulk's result (Dulk, 2000). It is important from our point of view that these bursts are associated with the same active region. This fact excludes the possible dispersion connected with the generation of radio emission in different plasmas. Nevertheless the calculated dependences are not as steep as those obtained for the Newkirk model. This can be interpreted as the acceleration of electrons with distance (Zhang, Wang, and Ye, 2018). Yet there is another explanation of this fact. Mann et al. (1999) showed that Newkirk model corresponds to a Boltzmann distribution for the plasma with temperature $T = 1.4 \cdot 10^6 K$. The drift rate for type III bursts generated in the plasma with a Boltzmann distribution $n(r) = n_s \cdot \exp[\frac{A}{R_S}(\frac{R_S}{r}-1)]$ (Mann et al., 1999) has the form

$$\frac{df}{dt} = \frac{f}{2}(-1)\frac{A}{r^2}V \approx -\frac{\mu G M_S \cdot f}{2k_B \cdot T \cdot r^2}V \qquad (2)$$

where $\mu$ is the mean molecular weight, $G$ is the gravitational constant, $M_\odot$ is the mass of the Sun, $k_B$ is the Boltzmann's constant, $T$ is the plasma temperature. So the drift rate increases with decreasing temperature. Thus the obtained tendency of increasing of drift rates means that the temperature of the coronal plasma decreases with distance. It looks rather reasonable from the physical point of view. Detailed analysis of such situation is beyond the scope of this article and will be considered in the new model of the solar corona with non constant temperature (Shergelashvili et al., 2020). Very important in this sense is the analysis not only of spectral characteristics of type III bursts but also of the spatial properties of the type III sources at different frequencies in a wide frequency band obtained from heliographic and interferometer observations. Nowadays such possibilities exist and first results have been obtained already by LOFAR (Mann et al., 2018) and UTR-2 (Melnik et al., 2017).

### 3.2. Durations of Type III and Type IIIb Bursts

Comparison of type III and type IIIb burst durations shows that those of the harmonic are 6 - 8 times longer than those of the fundamental both in the frequency band 10 - 30 MHz (Melnik et al., 2018) and also the wider frequency band of 10 - 70 MHz. Moreover functional dependences are also different in general. It is very strange because both fundamental and harmonic are generated by the same electron beams. It was shown in Melnik et al. (2017), that the duration of the harmonic was defined by the longitudinal size $L$ and the electron beam velocity $V$, as $D = L/V$. Then for type III burst on 6 April the longitudinal size of its source at 50 MHz was $L = VD \approx 6.3'$. This size is in a good agreement with the size of the type III burst source obtained from the interferometer observations by LOFAR (Mann et al., 2018). The size of the source for this burst at 16 MHz is twice larger. For type III bursts observed on 8 and 9 April the sizes of the sources are $8'$ at 30 MHz and $21'$ at 14 MHz and $9'$ at 30 MHz and $27'$ at 10 MHz respectively. We see that sizes of all discussed type III sources are quite similar. For these sizes the brightness temperatures for type III bursts increased from $10^9 K$ at 30 MHz to $10^{10} K$ at 10 MHz.

Durations of fundamental type IIIb bursts are significantly smaller than harmonic durations and as already noted (Melnik et al., 2014; Shevchuk et al., 2016) were approximately equal to the plasma collision time

$$\tau \approx \frac{T_e^{3/2} m^{1/2}}{4\pi e^4 n L_e} \qquad (3)$$

where $T_e$ is the electron temperature, $m$ is the electron mass, $e$ is the electron charge and $L_e$ is the Coulomb logarithm. If this is correct then $D \sim \tau \sim n^{-1} \sim f^{-2}$ resulting close to the duration of type IIIb burst on 6 and 9 April. Moreover the temperature of plasma should be $1.7 \cdot 10^6 K$ according to Equation 3 which appears to be quite reasonable.

It is necessary to note that durations of 30 type III bursts observed by PSP at the beginning of April 2019 at 10 MHz were about 10 - 20 s (Krupar et al., 2020). These values are close to those obtained in this article for type III bursts which were the harmonic. So we can conclude that PSP observed also the harmonic but not the fundamental component. This confirms the conclusion by Krupar et al.(2020). Furthermore Alvarez and Haddock (1973a) obtained the dependence for durations of type III burst in the frequency band of 50 kHz - 3.5 MHz as $D \sim 10^{7.7} f^{-0.95}$ (in our approximation $D \sim 4(f/30MHz)^{-0.95}$) which does not differ notably from the dependences in this article and from those by Krupar et al. (2020) $D \sim 62.72 f^{-0.6}$ (in our approximation $D \sim 8.1(f/30MHz)^{-0.6}$). Note that such large durations of harmonic type III bursts cannot be obtained by scattering on random density irregularities (Kontar et al., 2019; Krupar et al., 2020).

**4. Conclusion**

In this article we analyzed three events including type III and type IIIb bursts, which were observed on 6, 8, and 9 April 2019 by radio telescopes URAN-2 and GURT in the frequency band of 8 - 80 MHz. We showed that these bursts were the harmonic and fundamental correspondingly, taking into account their polarization (not higher than 20% for type III burst and as high as 80% for type IIIb bursts) and frequency ratios (from 1.78 to 2) (Suzuki and Dulk, 1985). It was shown that drift rates dependences on frequency were in good agreement with the assumption that electron beams with velocities 0.17-0.2 c propagate in a Newkirk solar corona. We attribute the large dispersion in drift rates of type III bursts observed by NDA (Zhang, Wang, and Ye, 2018) to the fact that analyzed bursts could be a mixture of both fundamental and harmonic and could be generated by electron beams with different velocities. It is worth mentioning that type III bursts were generated under different conditions. According to our observational data the obtained drift rate dependences are steeper than those for the Newkirk model. We assume that this can be related to decreasing temperature of the coronal plasma at distances 1.35 - 6.5 $R_\odot$.

Regarding durations of type III and type IIIb bursts in the frequency band of 8 - 80 MHz we confirm that in pairs IIIb-III they essentially differ, 6 - 8 times, at the same frequency. As a rule functional dependences of duration on frequency for type III and type IIIb bursts are different. Brightness temperatures of type III burst increase with decreasing frequency from $10^9 K$ at 30 MHz to $10^{10} K$ at 10 MHz. Durations of type IIIb bursts and actually durations of stria-bursts can be close to the collision time. The cause of why the fundamental and harmonic durations differ essentially, remains misunderstood. The fact that type IIIb bursts have a fine frequency structure and type III bursts do not in IIIb-III pairs in the decameter range, could be also connected with special features in the generation of these bursts. Though it should be noted that Loi, Cairns, and Li (2014) overcame this difficulty assuming that type III electron beams propagate through plasma with Kolmogorov inhomogenuities.

**Acknowledgments** The research was supported by projects Gorizont (0120U100234), Spectr-3 (0117U000245) and "Study of the polarization characteristics of decameter radio emission from space sources using the URAN-2 radio telescope" (0118U009760) of the National Academy of Sciences of Ukraine.

**Disclosure of Potential Conflicts of Interest** The authors declare that they have no conflicts of interest.